\documentclass{ws-ijmpa}
\usepackage{mathrsfs}

\begin{document}

\markboth{V.~I.~Kuksa, N.~I.~Volchanskiy} {Factorization effects
in a model of unstable particles}

%
\catchline{}{}{}{}{}
%

\title{FACTORIZATION EFFECTS IN A MODEL \\
OF UNSTABLE PARTICLES}

\author{V.~I.~KUKSA\footnote{kuksa@list.ru}}

\address{Research Institute of Physics, Southern Federal University, Rostov-on-Don 344090, Russia}

\author{N.~I.~VOLCHANSKIY\footnote{nikolay.volchanskiy@gmail.com}}

\address{Research Institute of Physics, Southern Federal University, Rostov-on-Don 344090, Russia}

\maketitle

\begin{history}
\received{Day Month Year} \revised{Day Month Year}
\end{history}

\begin{abstract}
The effects of factorization are considered within the framework of the model of unstable particles with a smeared mass. It is shown that two-particle cross section and three-particle decay width can be described by the universal factorized formulae for an unstable particles of an arbitrary spin in an intermediate state. The exact factorization is caused by the specific structure of the model unstable-particle propagators. This result is generalized to complicated scattering and decay-chain processes with unstable particles in intermediate states. We analyze applicability of the method and evaluate its accuracy.

\end{abstract}

\ccode{PACS number: 11.30.Pb}

\section{Introduction}

Unstable particles (UP's) are described usually by dressed propagator or with the help of the $S$-matrix with complex pole. The problems of these descriptions have been under considerable discussion for many decades
[\refcite{1}]--[\refcite{7}] (and references therein). There are
also other approaches such as time-asymmetric QFT of UP's
[\refcite{6}], effective theory [\refcite{8,9}], modified
perturbative theory approach [\refcite{10}], and phenomenological
QF model of UP's with smeared mass [\refcite{11,12}]. In this
work we consider some remarkable properties of the model [\refcite{11,12}] which
are caused by mass smearing and lead to the factorization effects
in the description of the processes with UP in an intermediate
state.

The model under consideration is based on the time-energy
uncertainty relation (UR). Despite their formal uniformity,
various UR's have different physical nature. This point has been
discussed for many years beginning with Heisenberg's formulating
the uncertainty principle (for instance, see
[\refcite{13}]--[\refcite{16}] and references therein). The first
model of UP based on the time-energy UR was suggested in
[\refcite{17}]. Time-dependent wave function of UP in its rest
frame was written in terms of the Fourier transform which may be
interpreted as a distribution of mass values, with a spread,
$\delta m$, related to the mean lifetime $\delta \tau =1/\Gamma$
by uncertainty relation [\refcite{14,17}]:
\begin{equation}\label{1.1}
\delta m\cdot\delta\tau\sim 1,\,\,\,\mbox{or}\,\,\,\delta m\sim
\Gamma\,\,\,(c=\hbar =1).
\end{equation}
Thus, from the time-energy UR for the unstable quantum system, we
are led to the concept of mass smearing for UP, which is described
by UR (\ref{1.1}). Implicitly (indirectly) the time-energy UR, or
instability, is usually taken into account by using the complex
pole in $S$-matrix or dressed propagator which describes UP in an
intermediate state. Explicit account of the relation (\ref{1.1})
is taken by describing UP in a final or initial states with the
help of the mass-smearing effect. From Eq. (\ref{1.1}) it follows
that this effect is noticeable if UP has a large width. Mass
smearing was considered in the various fields of particle
physics---in the decay processes of UP with large width
[\refcite{12}], in the boson-pair production [\refcite{18,19}],
and in the phenomenon of neutrino oscillations [\refcite{16,20}].
In these papers, the efficiency of the mass-smearing conception
was demonstrated in a wide class of processes.

The effects of exact factorization in the two-particle scattering
and three-particle decay were shown for the cases of scalar,
vector, and spinor UP in Refs.~[\refcite{21,22}]. Factorized
formulae for cross section and decay width were derived exactly
(without any approximations) for the tree-level processes in the
physical gauge. However, the results can be generalized taking
account of principal part of radiative corrections
[\refcite{12,18,19,21}] (see Section 3). The factorization method
that is based on the exact factorization in the simplest processes
was suggested in [\refcite{22a}]. In this work we generalize the
results of [\refcite{22a}] to the case of UP with spin $J=3/2$ and
apply them to some complicated processes. In the second section we
describe the main elements of the model that lead to the effects
of factorization. These effects in the case of the simplest
processes are considered in Section 3 for UP with
$J=0,\,1/2,\,1,\,3/2$. The factorization method based on the results
of the third section is applied to some more
complicated processes in Section 4. In this section, we also
consider the accuracy of the calculations. Some conclusions are
made concerning the applicability and advantages of the method in
the fifth section.

\section{Model of UP with a smeared mass}

In this section we present the elements of the model
[\refcite{11,12}] that are used directly in the factorization
method. The field function of the UP is a continuous superposition
of the standard ones defined at a fixed mass with a weight
function of the mass parameter $\omega(\mu)$ which describes mass
smearing. As a result of this, an amplitude of the process with
UP in a final or initial state has the form:
 \begin{equation}\label{2.1}
 A(k,\mu)=\omega(\mu)A^{st}(k,\mu),
 \end{equation}
where $A^{st}(k,\mu)$ is an amplitude defined in a standard way at
fixed mass parameter $\mu$, and $\omega(\mu)$ is a model weight
function.

The model Green's function has a convolution form with respect to
the mass parameter $\mu$ (the Lehmann representation). In the case
of scalar UP it is as follows:
 \begin{equation}\label{2.2}
 D(x)=\int D(x,\mu)\,\rho(\mu)\,d\mu,\,\,\,\rho(\mu)=|\omega(\mu)|^2,
 \end{equation}
where $D(x,\mu)$ is defined in a standard way for a fixed
$\mu = m^2$ and $\rho(\mu)$ is a probability density of the mass parameter
$\mu$.

The model propagators of scalar, vector, and spinor unstable
fields in momentum representation are given by the following
expressions [\refcite{12}]:
\begin{align}\label{2.3}
 D(q)=&i\int\frac{\rho(\mu)\,d\mu}{q^2-\mu+i\epsilon},\,\,\,D_{mn}(q)=-i\int\frac{g_{mn}-
 q_{m}q_{n}/\mu}{q^2-\mu+i\epsilon}\rho(\mu)\,d\mu,\notag\\
 &\hat{G}=i\int\frac{\hat{q}+\sqrt{\mu}}{q^2-\mu+i\epsilon}\rho(\mu)\,d\mu,
 \qquad q=\sqrt{(q_iq^i)}.
\end{align}
The model propagators are completely defined if the function $\rho(\mu)$ is
determined.

Here, we generalize the results of the works [\refcite{12,21,22}]
to include the case of unstable fields with spin $J=3/2$. The propagator of
this field is defined in [\refcite{23,24}] and smearing its mass, $M^2\to\mu$, gives:
\begin{align}\label{2.4}
    &\hat{G}_{mn}(q)=\int\rho(\mu)\,d\mu \{
    -\frac{\hat{q}+\sqrt{\mu}}{q^2-\mu+i\epsilon} ( g_{mn} -\frac{1}{3} \gamma_{m}
    \gamma_{n} - \frac{\gamma_{m} q_{n} -\gamma_{n} q_{m}}{3\sqrt{\mu}} - \frac{2}{3}
    \frac{q_{m} q_{n}}{\mu})\}.
\end{align}

Determination of the weight function $\omega(\mu)$ or
corresponding probability density $\rho(\mu)=|\omega(\mu)|^2$ can
be done with the help of the various methods [\refcite{12}]. Here
we consider the definition of $\rho(\mu)$ which leads to the
effect of exact factorization. We match the model propagator of
scalar UP to the standard dressed one:
\begin{equation}\label{2.5}
 \int\frac{\rho(\mu)d\mu}{k^2-\mu+i\epsilon}\longleftrightarrow
 \frac{1}{k^2-M^2_0-\Pi(k^2)}\,,
\end{equation}
where $\Pi(k^2)$ is a conventional polarization function. It was
shown in [\refcite{11,12}] that the correspondence (\ref{2.5})
leads to the definition:
\begin{equation}\label{2.6}
 \rho(\mu)=\frac{1}{\pi}\,\frac{Im\Pi(\mu)}{[\mu-M^2(\mu)]^2+[Im\Pi(\mu)]^2}\,,
\end{equation}
where $M^2(\mu)=M^2_0+Re\Pi(\mu)$. Substitution of the expression
(\ref{2.6}) into (\ref{2.3}) and integration over $\mu$ lead to
the results:
\begin{equation}\label{2.7}
 D_{mn}(q)=i\frac{-g_{mn}+q_m
 q_n/q^2}{q^2-M^2(q^2)-iIm\Pi(q^2)}
\end{equation}
and
\begin{equation}\label{2.8}
 \hat{G}(q)= i\frac{\hat{q}+q}{q^2-M^2(q^2)-iq\Sigma(q^2)}\,.
\end{equation}
In analogy with these definitions we get the expression for the
propagator of vector-spinor unstable field:
\begin{equation}\label{2.9}
    \hat{G}_{mn}(q)=
    -\frac{\hat{q}+q}{q^2-M^2(q^2)-iq\Sigma(q^2)} \{ g_{mn} -\frac{1}{3} \gamma_{m}
    \gamma_{n} - \frac{\gamma_{m} q_{n} -\gamma_{n} q_{m}}{3q} - \frac{2}{3}
    \frac{q_{m} q_{n}}{q^2}\}.
\end{equation}
Note that in Eqs. (\ref{2.8}) and (\ref{2.9}) we have substituted
$q\Sigma(q^2)$ for $\Pi(q^2)$. The expressions
(\ref{2.5})--(\ref{2.9}) define an effective theory of UP's which
follows from the model and the definition (\ref{2.6}), that is
from the correspondence (\ref{2.5}). In this theory the numerators
of the expressions (\ref{2.7})--(\ref{2.9}) differ from the
standard ones. The correspondence between standard and model
expressions for the cases of vector (in unitary gauge) and spinor
UP is given by the interchange $m \leftrightarrow q$ in the
numerators of the standard and model propagators (here
$q=\sqrt{q_i q^i})$. As a result the structures of the model
propagators lead to the effect of exact factorization, while the
standard ones lead to approximate factorization (see the next
section).

\section{Effects of factorization in the processes\\ with UP in an intermediate state}

Exact factorization is stipulated by the following properties of the model:\\
a) smearing the mass shell of UP in accordance with the time-energy UR;\\
b) specific structure of the numerators of the propagators, Eqs.~(\ref{2.7})--(\ref{2.9}).\\
The first factor allows us to describe UP in an intermediate state
with the momentum $q$ as a particle in a final or initial states
with the variable mass $m^2=q^2$. In this case, UP is described by
the following polarization matrices which differ from the standard
on-shell ones by the change $m\to q$ (see also [\refcite{12}]):
\begin{align}\label{3.1}
&\sum_{a=1}^{3} e^a_m(\vec{q})\dot{e}^a_n(\vec{q})=-g_{mn}+\frac{q_m q_n}{q^2}\,\,\,\,\,\mbox{(vector UP)};\notag\\
&\sum_{a=1}^{2}u^{a,\mp}_i(\vec{q})\bar{u}^{a,\pm}_k(\vec{q})=\frac{1}{2q^0}(\hat{q}+q)_{ik}
\,\,\,\,\,\mbox{(spinor UP)};\notag\\
&\hat{\Pi}_{mn}(q)=-\frac{1}{4} ( \hat{q} + q ) \left[g_{mn}
-\frac{1}{3} \gamma_{m} \gamma_{n}
    - \frac{\gamma_m q_n -\gamma_n q_m}{3q} - \frac{2}{3} \frac{q_m q_n}{q^2}
    \right],\notag\\
    &\mbox{(vector-spinor UP)}.
\end{align}
The second factor is the coincidence of the expressions for the
propagator numerators (\ref{2.7})--(\ref{2.9}) and for the
polarization matrices (\ref{3.1}). It allows us to represent the
amplitude of the process with UP in an intermediate state (see
Fig.1) in a partially factorized form:
\begin{equation}\label{3.2}
\mathit{M}(p,p',q)=K\sum_{a}\frac{\mathit{M}^{(a)}_1(p,q)\cdot
\mathit{M}^{(a)}_2(p',q)}{P(q^2,M^2)},
\end{equation}
where $\mathit{M}^{(a)}$ is a spiral amplitude. The representation
(\ref{3.2}) is a precondition of factorization, while full exact
factorization occurs in the transition probability. The effect of
factorization is illustrated in Fig.1 where UP in an intermediate
state is signed by crossed line.

\begin{figure}[!ht]
 \centerline{\epsfig{file=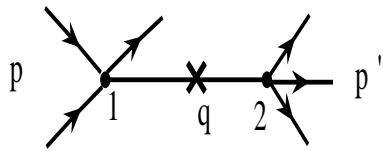,height=2cm,width=5cm}}
 \caption{Factorization in the reducible diagram.}
 \label{fig:Born1}
\end{figure}

Now, we demonstrate the factorization effect in the case of the
simplest basic elements of the tree processes, where UP is in the
$s-$ channel intermediate state. The vertices are described by the
simplest standard Lagrangians for scalar, vector, and spinor
particles (see Refs. [\refcite{21,22}]). Here we add the
Lagrangian that describes the interaction of vector-spinor
particles with lower-spin ones [\refcite{25}]--[\refcite{27}]:
\begin{align}\label{3.3}
    \mathscr{L}_{int}&=
    \frac{f}{m_\pi} \bar{\Psi}^\mu \bar{\Theta}_{\nu\mu} (z,\lambda) N \pi^{,\nu}_a +
    \mbox{h.c.}\notag\\
    &+\frac{g}{2 M_N} \bar{\Psi}^\mu \bar{\Theta}_{\nu\mu} (x,\lambda)
    \gamma^{\xi} \gamma_5 N\rho_{\xi}{}^{\nu} +
    \mbox{h.c.},
\end{align}
where
\begin{equation}\label{3.4}
    \Theta_{\mu\nu} (\lambda,\lambda') =
        g_{\mu\nu} + \frac{\lambda'-\lambda}{2\left( 2 \lambda - 1 \right)} \gamma_\mu
        \gamma_\nu\,,
\end{equation}
$x$, $z$ are off-shell parameters, and $\lambda$ is the parameter of
the Lagrangian of the Rarita-Schwinger free field.

Let us consider the simplest processes which are used further as the
basic elements of the method. Note that calculations are
made at the "tree level" in the framework of the effective
theory, which follows from the model with mass smearing and contains self-energy
corrections in the function $\rho(\mu)$. This effective theory is
not a gauge one, but the definitions of the model propagators
(\ref{2.3}) are given in analogy with the standard physical
gauge (unitary gauge for massive fields).

The first element is two-particle scattering with UP of any type
in the intermediate state.
\begin{figure}[!ht]
 \centerline{\epsfig{file=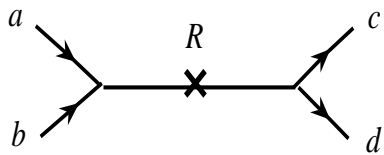,height=2cm,width=5cm}}
 \caption{\footnotesize Factorization in $2\to 2$ scattering diagram}
 \label{fig:Born2}\end{figure}

By straightforward calculation at the tree level it was shown that
the cross-section for all permissible combinations of particles
$(a,b,R,c,d)$ can be represented in the universal factorized form
[\refcite{21}]:
\begin{equation}\label{3.5}
 \sigma(ab\rightarrow R\rightarrow cd)=\frac{16\pi(2J_R+1)}
 {(2J_a+1)(2J_b+1)\bar{\lambda}^2(m_a,m_b;\sqrt{s})}
 \frac{\Gamma^{ab}_R(s)\Gamma^{cd}_R(s)}{|P_R(s)|^2}.
\end{equation}
Here, $s=(p_a+p_b)^2$, $P_R(s)$ is the denominator of the
propagator of an unstable particle $R(s)$ which is defined by
Eqs.~ (\ref{2.7}) and (\ref{2.8}), $\lambda(m_a,m_b;\sqrt{s})$ is
normalized K\"{a}ll\'{e}n function and
$\Gamma^{ab}_R(s)=\Gamma(R(s)\to ab)$ is a partial width of the
particle $R(s)$ with variable mass $m=\sqrt{s}$. Note that exact
factorization in the process under consideration always takes
place for scalar $R$ in both the standard and model treatment. In
the case of vector and spinor $R$ the factorization is exact in
the framework of the model only. In the standard treatment the
expression (\ref{3.5}) is valid in the narrow-width approximation
(see, for instance, Eq.~(37.51) and corresponding comment in
[\refcite{27a}]). If the diagram depicted in Fig.2 is included into a more
complicated one as a sub-diagram, where the particle $c$ and/or $d$
are unstable, then we have to generalize partial width, for
instance, $\Gamma^{cd(q)}_R (s,q)=\Gamma(R(s)\to cd(q))$, where
$d(q)$ is unstable particle $d$ with mass $m^2=q^2$ (see the next
section). Note that Eq.(\ref{3.5}) differs from the corresponding
equation (6) in [\refcite{21}] and equation (42) in [\refcite{22}]
which contain misprints. Correct expressions can be got in
[\refcite{21,22}] by modification $k_a k_b/k_R\to k_R/k_a k_b$,
where $k_p=2J_p+1$ and $J_p$ is spin of the particle $p =
a,\,b,\,R$.

It was shown in Ref.[\refcite{21}] that Eq.(\ref{3.5}) is valid
for the cases of scalar ($J=0$), vector ($J=1$), and spinor
($J=1/2$) unstable particles. In this work we check by direct
calculation that Eq.(\ref{3.5}) is also correct in the case of UP
with $J=3/2$, when Eq.~(\ref{2.9}) is used (for instance,
$\Delta$-resonance production). We should note that the
expressions which involve vector-spinor UP are valid for the
particles on the mass shell in the framework of the standard
treatment. However, in the framework of the model, UP is always on
its smeared mass shell and these expressions are valid in general
case. Moreover, the part of expressions which contains off-shell
parameters as well as $\lambda$-parameter disappears in widths and
cross-sections. Hence the condition of factorization---the
coincidence of the polarization matrix and the numerators of the
propagators---is fulfilled in the case under consideration (see
Eqs. (\ref{2.9}) and (\ref{3.1})).

\begin{figure}[!ht]
 \centerline{\epsfig{file=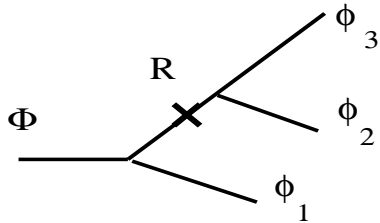,height=3cm,width=5cm}}
 \caption{\footnotesize Factorization in $1\to 3$ decay diagram}
 \label{fig:Born3}\end{figure}

The second basic element is a three-particle decay with UP in the
intermediate state $\Phi\to\phi_1R\to\phi_1\phi_2\phi_3$ (Fig.\ref{fig:Born3}), where
$R$ is UP of any kind.

By straightforward calculations it was shown that the
three-particle partial width at the tree level can be represented
in the universal factorized form [\refcite{22}]:
\begin{equation}\label{3.6}
  \Gamma(\Phi\rightarrow\phi_1 \phi_2 \phi_3)=\int_{q^2_1}^{q^2_2}\Gamma(\Phi
  \rightarrow\phi_1 R(q))\,\frac{q\,\Gamma(R(q)\rightarrow\phi_2\phi_3)}
  {\pi \vert P_{R}(q)\vert ^2}\,dq^2\,,
\end{equation}
where $R$ is a scalar, vector or spinor UP, $q_1=m_2+m_3$ and
$q_2=m_{\Phi}-m_1$. By direct calculations we also check the
validity of the expression (\ref{3.6}) for the case of the UP with
$J=3/2$ (see the remark to Eq.(\ref{3.5})). It is seen clearly that
the formula (\ref{3.6}) can include any factorizable corrections.

By summing over decay channels of $R$, from
Eq.(\ref{3.6}) we get the well-known convolution formula for the
decays with UP in a final state [\refcite{22}]:
\begin{equation}\label{3.7}
 \Gamma(\Phi\rightarrow\phi_1 R) = \int_{q^2_1}^{q^2_2} \Gamma(\Phi\rightarrow
 \phi_1 R(q))\,\rho_{R}(q)\,dq^2\,.
\end{equation}
In Eq.(\ref{3.7}) smearing the mass of unstable state $R$ is
described by the probability density $\rho_R(q)$:
\begin{equation}\label{3.8}
 \rho_{R}(q)=\frac{q\,\Gamma^{tot}_R(q)}{\pi\,\vert P_{R}(q)\vert^2}.
\end{equation}
The expression (\ref{3.8}) is connected with Eq.(\ref{2.6}) by the
relation $Im\Pi(q)=q\, \Gamma^{tot}_R(q)$.

The factorized expressions (\ref{3.6}) and (\ref{3.7}) are
applied successfully for the description of the decay $B\to \rho D$
[\refcite{12}], decay properties of $\phi(1020)$-meson
[\refcite{12}] and $t$-quark [\refcite{31,32,33}], lightest
chargino and next-to-lightest neutralino [\refcite{33a}].
Moreover, the formula (\ref{3.6}) describes
the decays $\mu\to e\bar{\nu}_e \nu_{\mu}$, $\tau\to
e\bar{\nu}_e\nu_{\tau}$, and $\tau^-\to \nu_{\tau}\pi^-\pi^0$ with great accuracy (see
the next section). It should be noted that, in analogy with two-particle
scattering, the factorization in the expression for the width
(\ref{3.6}) is also exact within the framework of the model and
approximate in the standard treatment (convolution method). Besides,
we note that the expressions (\ref{3.5}) and (\ref{3.6})
significantly simplify calculations in comparison with the standard
ones.

\section{Factorization method in the model of UP's\\ with smeared mass}

\begin{figure}[!ht]
 \centerline{\epsfig{file=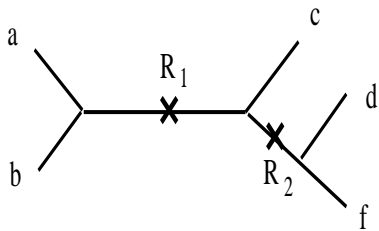,height=3cm,width=5cm}}
 \caption{\footnotesize Factorization in $2\to 3$ scattering-decay diagram}
 \label{fig:Born4}\end{figure}

The method is based on exact factorization of the simplest
processes with UP in an intermediate state that were considered in
Section 3. The factorization method has applicability to such Feynman diagrams that can be disconnected into two components by cutting some line corresponding to timelike momentum transfer. For instance, it is applicable to the
complicated scattering and decay-chain processes which can be
reduced to a chain of the basic elements (\ref{3.5}) and
(\ref{3.6}). Next, we consider some examples of such processes.

\begin{equation}\label{4.1}
1)\, a+b\to R_1\to c+R_2\to c+d+f \text{ (Fig.\ref{fig:Born4})}.
\end{equation}

The cross-section of this process is a combination of the
expressions (\ref{3.5}) and (\ref{3.6}):
\begin{align}\label{4.2}
 &\sigma(ab\to R_1\to cdf)=\notag\\
 &\frac{16k_{R_1}}{k_a k_b\bar{\lambda}^2(m_a,m_b;\sqrt{s})}\frac{\Gamma^{ab}_{R_1}(s)}{|P_{R_1}(s)|^2}
 \int_{q^2_1}^{q^2_2}\Gamma(R_1(s)\to
 c R_2(q))\,\frac{q\,\Gamma^{df}_{R_2}(q)}{|P_{R_2}(q)|^2}\,dq^2,
\end{align}
where $k_p=2J_p+1$. It should be noted that the factorization
effectively reduces the number of independent kinematic
variables which have to be integrated. In the standard approach
for the process $2\to 3$ the number of the variables, which
uniquely specify a point in the phase space, in the general case
is $N=3n-4=5$, from which four variables have to be integrated
[\refcite{28}]. Some of this variables can be integrated out if a
specific symmetry of the process occur. In the framework of the
approach suggested the number of integrated variables is always
$N_M=1$. The same effect of variable reduction takes place for the
case of the basic processes of scattering and decay which were
considered in the previous section.

\begin{figure}[!ht]
 \centerline{\epsfig{file=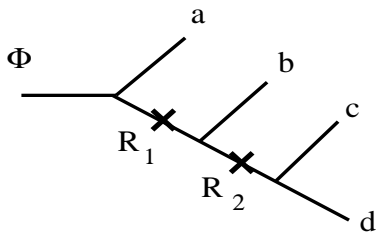,height=3cm,width=5cm}}
 \caption{\footnotesize Factorization in $1\to 4$ decay diagram}
 \label{fig:Born5}\end{figure}

The expression (\ref{4.2}) can be used for fast evaluation
of cross sections of some scattering processes both in cosmology
and collider physics. For instance, it is valid for the
description of the annihilation process with the lightest
supersymmetric particle in a final state [\refcite{33a}].

\begin{equation}\label{4.3}
2)\, \Phi\to a+R_1\to a+b+R_2\to a+b+c+d\text{ (Fig.\ref{fig:Born5})}.
\end{equation}

\begin{figure}[!ht]
 \centerline{\epsfig{file=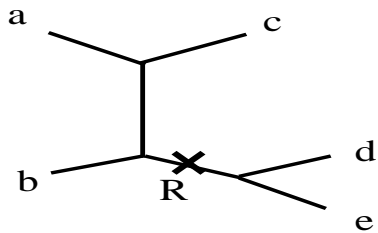,height=3cm,width=5cm}}
 \caption{Factorization in $a+b\to c+R\to c+d+e$ process.}
 \label{fig:Born6}
\end{figure}

The width of this decay-chain process is given by doubling the
formula (\ref{3.6}):
\begin{align}\label{4.4}
\Gamma(\Phi\to abcd)&=\frac{1}{\pi^2}\int_
{q^2_1}^{q^2_2}\frac{q\,\Gamma(\Phi
  \rightarrow a R_1(q))}{\vert P_{R_1}(q)\vert ^2}\times\notag\\&\times\int_{g^2_1}^{g^2_2}\Gamma(R_1(q)\to bR_2(g))
  \,\frac{g\,\Gamma(R_2(g)\rightarrow cd)}
  {\vert P_{R_2}(g)\vert ^2}\,dg^2\,dq^2.
\end{align}
Note that in the general case of $n$-particle decay the number of
kinematic variables, which uniquely specify a point in the phase
space, is $N=3n-7=5$ [\refcite{28}], while the
method gives $N_M=2$ (see comment to the previous case). Thus, we
have a significant simplification of calculations. Some
examples of processes which can be described by the compact
formula (\ref{4.4}) are considered in [\refcite{33}], where the
relation between convolution method and decay-chain method is
analyzed.

\begin{equation}\label{4.5}
3)\, a+b\to c+R\to c+d+e\text{ (Fig.\ref{fig:Born6})}.
\end{equation}

\begin{figure}[!ht]
 \centerline{\epsfig{file=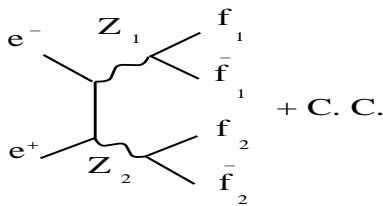,height=2cm,width=5cm}}
 \caption{$Z$-pair production process.}
 \label{fig:Born7}
\end{figure}

The cross-section of this $t$-channel process is described by
convolution of the cross-section $\sigma(ab\to cR)$ and the width
$\Gamma(R\to de)$:
\begin{equation}\label{4.6}
\sigma(ab\to cde)=\frac{1}{\pi}\int_{q_1^2}^{q_2^2}\sigma(ab\to
cR(q))\frac{q\Gamma(R(q)\to de)}{\vert P_{R}(g)\vert ^2}\,dq^2.
\end{equation}
This formula can be applied to the description of the processes
$e^+e^-\to \gamma Z\to \gamma f\bar{f}$ and $eN\to e\Delta \to
e\pi N$. The diagram in Fig.6 illustrates a class of processes at
the tree level with fermion-antifermion pair in the one-pole
approximation, i.e. generated from the decay of $R$ only. However,
the formula (\ref{4.6}) can be easily generalized taking account
of factorizable radiative corrections. For instance, such a
generalization of the expression (\ref{4.6}) was made in
[\refcite{28a}] for the description of the process $e^+e^-\to
\gamma Z\to \gamma \sum_f \nu_f\bar\nu_f$, $f=e,\,\mu,\,\tau$, with $\nu\bar\nu$-pairs being produced
by $Z$-decay only (``single-pole'' resonant production). Note that additional non-resonant ``ladder'' diagrams also contribute to the process $e^+e^-\to
\gamma \nu\bar\nu$. The resonant events can be still separated in certain kinematic regions [\refcite{28aa}]. In our calculations [\refcite{28a}] (with the kinematic cut corresponding to the event selection of Ref.[\refcite{28aa}]) we have taken into account ISR and principal part
of radiative corrections. These corrections do not change the
structure of the expression (\ref{4.6}) and satisfy the condition
of factorization. The results are in good agreement with the
experimental data and SM predictions at
$\sqrt{s}=185 - 210\,\mbox{GeV}$ [\refcite{28a}]. We should note,
however, that our calculations are valid in the energy domain, where
the selection of the resonant events is possible.

\begin{equation}\label{4.7a}
4)\, e^+e^-\to ZZ\to \sum_{i,k} \bar{f}_i f_i \bar{f}_k f_k\text{
(Fig.\ref{fig:Born7})}.
\end{equation}

Now, we consider the process of $Z$-pair production (or
four-fermion production in the double-pole approximation). Direct
application of the model to the process $e^+e^-\to ZZ$ or using
the factorization method for the full process $e^+e^-\to
ZZ\to \sum_{i,k} \bar{f}_i f_i \bar{f}_k f_k$ (double-pole
approach) gives the following expression for cross-section at the
tree level [\refcite{18}]:
\begin{equation}\label{4.8a}
\sigma^{tr}(e^+e^-\to ZZ)=\int\int\sigma^{tr}(e^+e^-\to
Z_1(m_1)Z_2(m_2))\,\rho_Z(m_1)\,\rho_Z(m_2)\,dm_1\,dm_2,
\end{equation}
where $\sigma^{tr}(e^+e^-\to Z_1(m_1)Z_2(m_2))$ is defined in
a standard way for the case of fixed boson masses $m_1$ and $m_2$,
and probability density of mass $\rho(m)$ is defined by the
expression:
\begin{equation}\label{4.7}
\rho_Z(m)=\frac{1}{\pi}\,\frac{m\,\Gamma^{tot}_Z(m)}{(m^2-M^2_Z)^2+(m\,\Gamma^{tot}_Z(m))^2}.
\end{equation}
Similar expressions can be written for the processes $e^+e^-\to
W^+W^-$ [\refcite{19}] and $e^+e^-\to ZH$ [\refcite{28a}]. To
describe exclusive processes, such as $e^+e^-\to ZZ\to
f_i\bar{f}_i f_k\bar{f}_k$, one has to substitute partial
$q$-dependent width $\Gamma_Z^i(q)=\Gamma(Z(q)\to f_i\bar{f}_i)$
into the expression (\ref{4.7}) instead of the total width
$\Gamma^{tot}_Z(m)$. Note that the expression similar to
(\ref{4.6}) can be written in the standard approach as a result of
integration over the phase space variables which describe
$4f$-states in the semi-analytical approximation (SAA)
[\refcite{28b}]. Within the framework of the model considered,
formula (\ref{4.6}) is derived exactly without any approximations.
Note also that as a rule the deviation of the standard exact
results from the model ones is negligible (see Eq. (\ref{4.12})).

The processes $e^+e^-\to ZZ, W^+W^-, \gamma Z, ZH$ were considered
in detail [\refcite{17,18,28a}] taking account
of the relevant radiative corrections. It was shown that the
results of the model calculations are in good agreement with the
experimental LEP II data and coincide with standard Monte-Carlo
results with great accuracy. At the same time the factorization
method significantly simplifies the calculation procedures in
comparison with the standard ones that should consider about ten thousands
diagrams (see, for example, [\refcite{28c}]
and comments in [\refcite{18}]).

Using the factorization method one can describe complicated decay-chain and scattering
processes in a simple way. The same results can occur within the
frame of standard treatment as the approximations. Such
approximations are known as narrow-width approximation (NWA)
[\refcite{29,30}], convolution method (CM) [\refcite{31}]--[\refcite{33}], decay-chain method (DCM) [\refcite{33}] and
semi-analytical approach (SAA) [\refcite{28b}]. All these
approximations get a strict analytical formulation within the
framework of the factorization method. For instance, NWA includes five assumptions which
was considered in detail in [\refcite{29}]. The factorization method
contains just one assumption---non-factorizable corrections are small
(the fifth assumption of NWA). The method suggested can be applied
to very complicated decay-chain and scattering processes by combining the expressions considered above. In this case
we have not a strict and general standard analog of such
approximation.

Now, we consider some ways of evaluation of the method error which
we define as the deviation of the model results from the strict
standard ones. For a scalar UP the error always equals zero in
accordance with the definition (\ref{2.5}). For a vector UP the
error is caused by the following difference:
\begin{equation}\label{4.8}
\delta\eta_{\mu\nu}=\eta_{\mu\nu}(q^2)-\eta_{\mu\nu}(m^2)=q_{\mu}q_{\nu}\frac{m^2-q^2}{m^2q^2},
\end{equation}
where $\eta_{\mu\nu}(m^2)$ and $\eta_{\mu\nu}(q^2)$ are standard
and model numerators of vector propagators in the physical gauge. In
the case of meson-pair production $e^+e^-\to
\rho^0,\omega,...\to\pi^+\pi^-, K^+K^-,\rho^+\rho^-,...$ the
deviation equals zero too, due to vanishing contribution of the
transverse parts of the amplitudes in both cases:
\begin{equation}\label{4.9}
\mathit{M}^{trans}(q)\sim
\bar{e}^-(p_1)\hat{q}e^-(p_2)=\bar{e}^-(p_1)(\hat{p}_1+\hat{p}_2)e^-(p_2)=0.
\end{equation}
In the case of the high-energy collisions $e^+e^-\to Z\to
f\bar{f}$ (we neglect $\gamma-Z$ interference) the transverse part
of the amplitude is:
\begin{equation}\label{4.10}
\mathit{M}^{trans}(q)\sim
\bar{e}^-(p_1)\hat{q}(c_e-\gamma_5)e^-(p_2)\cdot
\bar{f}^+(k_1)(c_f-\gamma_5)f^+(k_2)
\end{equation}
and we get at $q^2\approx M^2_Z$:
\begin{equation}\label{4.11}
\delta \mathit{M}\sim \frac{m_em_f}{M^2_Z}\frac{M_Z-q}{M_Z}.
\end{equation}
Thus, an error of the factorization method at the vicinity of resonance is always small,
moreover, it is suppressed by small factor $m_e m_f/M^2_Z$. The
similar estimations can be easily done for the case of a spinor
UP.

The relative deviation of the model cross section of the
boson-pair production with consequent decay of the bosons to
fermion pairs is [\refcite{18}]:
\begin{equation}\label{4.12}
\epsilon_f \sim
4\frac{m_f}{M}[1-M\int_{m^2_f}^{s}\frac{\rho(q^2)}{q}\,dq^2],
\end{equation}
where $M$ is a boson mass. For the case $f=\tau$ a deviation is
maximal, $\epsilon_{\tau}\sim 10^{-3}$. It should be noted that
the deviations that are caused by the approach at the tree level
are significantly smaller then the errors caused by the
uncertainty in taking account of radiative corrections
[\refcite{19}]. Thus, the the error of the method at the tree
level in the case of vector UP is, as a rule, negligible.

By straightforward calculations we evaluate the relative
deviations of the model partial width from standard one, that is,
$\epsilon =(\Gamma^M-\Gamma^{st})/\Gamma_M$ for the case of $\mu$
and $\tau$ decays. We get:
\begin{equation}\label{4.13}
\epsilon(\mu\to e\nu\bar{\nu})\approx 5\cdot
10^{-4};\,\,\,\epsilon(\tau\to e\nu\bar{\nu})\approx 3\cdot
10^{-6};\,\,\,\epsilon(\tau\to \mu\nu\bar{\nu})\approx 3\cdot
10^{-2}.
\end{equation}
The deviation is suppressed by the factor $k=m^2_{l1}/m^2_{l2}$,
which is small in the first and second case and large in the last
case. In the case of the decay $\tau^-\to \nu_{\tau}\pi^-\pi^0$,
suppression factor is very small,
$k=(m^2_{\pi_0}-m^2_{\pi_-})^2/m^4_{\tau} \sim 10^{-7}$.

In the case of a spinor UP in an intermediate state a deviation is
of the order of $(M_f-q)/M_f$. It can be large when $q$ is far
from the resonance region. However, in analogy with vector UP,
this deviation can be suppressed by small factor too, and we have
to control this effect in every case under consideration. The same
effect can occur in the case of the vector-spinor UP.

\section{Conclusion}

The model of UP's leads to effective theory of UP's with a
specific structure of vector, spinor and vector-spinor
propagators. Such a structure gives rise to the effects of exact
factorization in a broad class of the processes with UP's in
intermediate states. These effects allow us to develop the
factorization method for the description of the complicated
processes with participation of an arbitrary type of UP's. The
method suggested is simple and convenient tool for deriving the
formulae for cross sections and decay rates in the case of
complicated scattering and decay-chain processes. The
factorization method can be used as some analytical analog of NWA,
which enables us to evaluate the error of the approach at the tree
level in a simple way. We have shown that these errors, as a rule,
are significantly smaller then the ones caused by the uncertainty
in taking account of radiative corrections. It should be noted that the applicability of the method is limited to the energy scales where the non-resonant or non-factorizable contributions can be neglected.

The factorization method based on the model of UP with a smeared
mass can be treated in two various ways. On the one hand, it
follows from the specific structure of propagators and can be
interpreted as some heuristic (irrespective of the model) way to
evaluate decay rates and cross sections easily with the help of
the concise and convenient expressions. On the other hand, the
model, from which the factorization method follows, is based on the fundamental
properties of UP---time-energy UR (i.e., smearing the mass). Thus,
the method can be also used as some physical basis for development
of precision tools of rapid and easy calculations.

\end{document}